\documentstyle[aps,preprint]{revtex}
\newcommand{\ds}{\displaystyle}
\newcommand{\ltwid}{\raise.3ex\hbox{$<$\kern-.75em\lower1ex\hbox{$\sim$}}}
\newcommand{\rtwid}{\raise.3ex\hbox{$>$\kern-.75em\lower1ex\hbox{$\sim$}}}
\renewcommand{\theequation}{\arabic{section}.\arabic{equation}}

\title{Pseudogap phase formation in the crossover from
       Bose--Einstein condensation to BCS superconductivity}

\author{V.~P.~Gusynin$^{1}$\cite{email1},
        V.~M.~Loktev$^{1}$\cite{email2}
        and S.~G.~Sharapov$^{2}$\cite{perm}}

\address{$^{1}$Bogolyubov Institute for
               Theoretical Physics,\\
               252143 Kiev, Ukraine \\
         $^{2}$Department of Physics,
               University of Pretoria,\\
               0002 Pretoria, South Africa
}

\date{June 3, 1998}

\draft

\begin{document}
\tighten

\maketitle

\begin{abstract}
A phase diagram for a 2D metal with variable carrier density has been
derived. It consists of a normal phase, where the order parameter is
absent; a so-called ``abnormal normal'' phase where this parameter
is also absent but the mean number of composite bosons (bound pairs)
exceeds the mean number of free fermions; a pseudogap phase where the
absolute value of the order parameter gradually increases but its
phase is a random value, and finally a superconducting (here
Berezinski\u{i}--Kosterlitz--Thouless) phase.  The characteristic
transition temperatures between these phases are found.  The chemical
potential and paramagnetic susceptibility behavior as functions of
the fermion density and the temperature are also studied.  An attempt
is made to qualitatively compare the resulting phase diagram with the
features of underdoped high-$T_{c}$ superconducting compounds above
their critical temperature.
\end{abstract}
\section{Introduction}
\setcounter{equation}{0}

The study of the crossover region between superconductivity of Cooper
pairs and superfluidity of composite bosons is attracting much attention
due to its close relationship to the problem of describing
high-temperature superconductors (HTSC) (see, e.g., Refs.
\cite{Loktev.review,Randeria,Uemura}).  At present this region is
understood for 3D systems, both at zero and finite temperatures
\cite{Randeria.book,Haussmann}. The crossover in quasi-2D systems has also
been studied \cite{GLSh.PhysicaC}, albeit only partially, whereas for 2D
systems only the case of $T=0$  has been studied thoroughly
\cite{Randeria.book,GGL}.  This is related to the fact that fluctuations of
the charged complex order parameter in 2D systems are so large that they
destroy long-range order at any finite temperature
(Coleman--Mermin--Wagner--Hohenberg (CMWH) theorem \cite{Coleman}).
In this case the appearance of an inhomogeneous condensate with a
power--law decay for the correlations (the so--called
Berezinski\u{i}--Kosterlitz--Thouless (BKT) phase) is possible.  However
an adequate mathematical description for BKT phase formation is still
lacking.

Most previous analyses \cite{Varma,Serene,Tokumitu}
of the behavior of 2D systems at $T \neq 0$ have been based on
the Nozi\`eres--Schmitt--Rink approach \cite{Nozieres}. This approach
is simply a Gaussian approximation to the functional integral, and this
perhaps explains the difficulties faced in these calculations. On the
one hand, Gaussian fluctuations destroy long-range order in 2D, and if
one searches for the $T_{c}^{2D}$ at which order sets in, one
should obtain zero in accordance with the aforementioned theorems
\cite{Coleman}. On the other hand, taking Gaussian fluctuations into
account is completely inadequate to describe the BKT transition
\cite{Traven}.

Nonetheless, there has been some progress.  For example, the BKT
transition has been studied in relativistic $2+1$-theory
\cite{MacKenzie}, and the crossover from superconductivity to superfluidity
has been considered \cite{Drechsler}  as a function of the carrier density
$n_{f}$ (see also Ref. \cite{Zwerger}). However, the method employed in
Ref. \cite{Drechsler} to obtain the temperature $T_{\rm BKT}$ has several
drawbacks.  Most importantly, the equation for $T_{\rm BKT}$ was
obtained without considering the existence of a neutral (real) order
parameter $\rho$, whose appearance at finite $T$  does not violate the CMWH
theorem.

As we show below, $\rho$ defines the modulus of a multivalued complex
order parameter $\Phi$ for a 2D system.
As a result of allowing for a neutral order parameter, a region
where $\rho$ decays gradually to zero appears in the phase diagram of the
system. This region separates the standard normal phase with $\rho = 0$
from the BKT phase, where the correlations exhibit power-law decay. Despite
the exponential decay of correlations, this new region of states may be
expected to possess unusual properties, since $\rho$ plays the same role as
the energy gap $\Delta$ in the theory of ordinary superconductors in many
cases.\footnote{
To calculate the observed single-particle spectrum, of
course, carrier losses due to scattering of carriers by fluctuations of
the phase of the order parameter (and in real systems by dopants)
must be taken into account; see Ref. \cite{Loktev.Pogorelov}.}
The possible existence of such a phase, which in some sense is also normal,
may shed light on the anomalous behavior of the normal state of HTSC (see,
for example, the reviews in Refs. \cite{Loktev.review,Randeria} and
\cite{Pines.review}).  In particular, the
temperature dependencies of the spin susceptibility, resistivity, specific
heat, photoemission spectra, and other quantities
\cite{Randeria,Randeria.Nature} can be explained by the formation of either
a pseudogap or a spin gap in the region $T > T_{c}$.

Using a very simple continuum 2D model, this approach was first attempted
in a brief note \cite{JETP.Letters}, where we  calculated $T_{\rm BKT}$ and
$T_{\rho}$ ($T_{\rho}$ is the temperature defined by the condition $\rho= 0$)
self-consistently as functions of $n_{f}$, and established the boundaries of
this new {\em pseudogap} region, which lies between $T_{\rm BKT}$ and
$T_{\rho}$.

The purpose of this article is to develop this approach further. Using
the static paramagnetic susceptibility as an example,
we demonstrate that the pseudogap opens below $T_{\rho}$.
Furthermore, we analyze the difference between the commonly used
(see Refs. \cite{Uemura} and \cite{Randeria.book}) pairing temperature
$T_{P}$ and the temperature $T_{\rho}$ introduced here.
These temperatures turn out to be different if the chemical
potential $\mu < 0$. We also introduce here  an {\em abnormal normal}
phase, which lies between $T_{P}$ and $T_{\rho}$, where preformed bosons
exist. This more detailed study helps to clarify the physical import of
$T_{\rho}$, as well as the nature of the transition at $T_{\rho}$.
It was believed in the related model \cite{MacKenzie} that this is
a second-order phase transition. We argue however, that fluctuations in
the phase of the order parameter can transform the transition to a
crossover, as observed experimentally.

In Sec.~II  we present the model and the relevant formalism.
The equations for $T_{\rm BKT}$, $\rho$, $T_{\rho}$, and the chemical
potentials $\mu(T_{\rm BKT})$ and $\mu(T_{\rho})$ are derived in Sec.~III.
Since the technique employed to obtain the equation for
$T_{\rm BKT}$ is not widely used, we consider it useful to present a
detailed derivation of this equation. (The details of the calculation of
the effective potential and useful series are given in Appendix~A.)
The systems of equations for $T_{\rm BKT}$, $\rho(T_{\rm BKT})$,
$\mu(T_{\rm BKT})$ and $T_{\rho}$, $\mu(T_{\rho})$ are analyzed in Sec.~IV.
The difference between pairing temperature $T_{P}$ and the temperature
$T_{\rho}$ is discussed in Sec.~V.  Also discussed is the physical import
of $T_{\rho}$. Using the example of the static spin susceptibility, it is
shown in Sec.~VI that the resulting pseudogap phase can in fact be used to
explain the aforementioned anomalous properties of HTSC.

%%%%%%%%%%%%%%%%%%%%%%%%%%%%%%%%%%%%%%%%%%%%%%%%%%%%%%%%%%%%%%%%%%%%%
\section{Theoretical Framework}
%%%%%%%%%%%%%%%%%%%%%%%%%%%%%%%%%%%%%%%%%%%%%%%%%%%%%%%%%%%%%%%%%%%%%
\setcounter{equation}{0}

The simplest model Hamiltonian density for fermions confined to a 2D volume
$v$ is \cite{Randeria.book,GGL,Varma}
\begin{equation} {\cal H} =
\psi_{\sigma}^{\dagger}(x) \left(- \frac{\nabla^{2}}{2 m} - \mu \right)
    \psi_{\sigma}(x) - V \psi_{\uparrow}^{\dagger}(x)
       \psi_{\downarrow}^{\dagger}(x) \psi_{\downarrow}(x)
       \psi_{\uparrow}(x),
\label{Hamilton}
\end{equation}
where $x \equiv  \mbox{\bf r}, \tau$; $\psi_{\sigma}(x)$ is
a fermion field, $m$ is the effective fermion mass,
$\mu$ is the chemical potential, and $V$ is an effective local attraction
constant; we take $\hbar = k_{B} = 1$.

The Hubbard--Stratonovich method, which is standard for these
problems \cite{Kleinert}, can be applied to write the partition function
$Z({v}, \mu, T)$ as a functional integral over Fermi fields (Nambu
spinors) and the auxiliary field $\Phi = V
\psi_{\uparrow}^{\dagger} \psi_{\downarrow}^{\dagger}$. In contrast to
the usual method for calculating $Z$ in $\Phi$,
$\Phi^{\ast}$ variables, the parametrization $\Phi(x) = \rho(x)
\exp{[-i \theta(x)]}$ is more appropriate for presenting the
corresponding integral in two dimensions \cite{Witten} (see also
Refs. \cite{Thouless} and \cite{Beck}). When this replacement by
modulus--phase variables is implemented, it is evident that one must
also replace $\psi_{\sigma}(x) = \chi_{\sigma}(x) \exp{[i
\theta(x)/2]}$. Physically, this amounts to replacing the charged fermion
$\psi_\sigma(x)$ with a neutral fermion $\chi_\sigma(x)$ and spinless
charged boson $e^{i\theta(x)/2}$. Note that
while one may formally use any self-consistent definition of the
new variables, the physical condition that the macroscopic variable
$\Phi (x)$ be single-valued under $2 \pi$ rotations fixes the
parametrization. This was not taken into account in Ref.
\cite{JETP.Letters}, where a different parametrization was used.

As a result, one obtains
\begin{equation}
Z(v, \mu, T) = \int \rho\, {\cal D} \rho\, {\cal D} \theta
\exp{[-\beta \Omega (v, \mu, T, \rho(x), \partial \theta (x))]},
                 \label{partition.2D.effective}
\end{equation}
where
\begin{equation}
\beta \Omega (v, \mu, T, \rho (x), \partial \theta (x)) =
\frac{1}{V} \int_{0}^{\beta} d \tau \int d \mbox{\bf r}\,
\rho^{2}(x) - \mbox{Tr}\ln G^{-1} + \mbox{Tr} \ln G_{0}^{-1}
\label{Effective.Action.2D}
\end{equation}
is  the one-loop effective action, which depends on the modulus--phase
variables. The action (\ref{Effective.Action.2D}) can be expressed in terms
of the Green function of the initial (charged) fermions, which in the new
variables has the operator form
\begin{eqnarray}
G^{-1} & = & -\hat{I} \partial_{\tau} +
\tau_{3} \left(\frac{\nabla^{2}}{2 m} + \mu \right) +
\tau_{1} \rho(\tau, \mbox{\bf r})
\nonumber \\
& - &
\tau_{3} \left[
\frac{i \partial_{\tau} \theta(\tau, \mbox{\bf r})}{2} +
\frac{(\nabla \theta(\tau, \mbox{\bf r}))^{2}}{8 m} \right] +
\hat{I} \left[\frac{i \nabla^{2} \theta(\tau, \mbox{\bf r})}{4 m} +
\frac{i \nabla \theta(\tau, \mbox{\bf r}) \nabla}{2 m} \right].
                          \label{Green.fermion.phase}
\end{eqnarray}
The free fermion Green function $G_{0} = G|_{\mu, \rho, \theta=0}$
provides a convenient regularization in the process of calculation.
It is important that neither the smallness nor slowness of the variation
of the phase of the order parameter is assumed in obtaining expression
(\ref{Effective.Action.2D}). In other words, it is formally exact.

Since the low-energy dynamics of phases for which
$\rho \neq 0$ is governed mainly by long-wavelength fluctuations
of $\theta(x)$, only the lowest-order derivatives of the phase need be
retained in the expansion of
$\Omega(v, \mu, T, \rho(x), \partial \theta(x))$:
\begin{equation}
\Omega (v, \mu, \rho(x), \partial \theta(x))  \simeq
\Omega _{\rm kin} (v, \mu, T, \rho, \partial \theta(x)) +
\Omega _{\rm pot} (v, \mu, T, \rho),
\label{kinetic.phase+potential}
\end{equation}
where
\begin{equation}
\Omega _{\rm kin} (v, \mu, T, \rho, \partial \theta(x))
 = T\, \mbox{Tr} \sum_{n=1}^{\infty}
\left. \frac{1}{n} ({\cal G} \Sigma)^{n} \right|_{\rho = const}
\label{Omega.Kinetic.phase}
\end{equation}
and
\begin{equation}
\Omega _{\rm pot} (v, \mu, T, \rho)  =
\left. \left(\frac{1}{V} \int d \mbox{\bf r} \rho^{2} -
 T\, \mbox{Tr}\ln{\cal G}^{-1} +
 T\, \mbox{Tr}\ln G_{0}^{-1} \right) \right|_{\rho = const}.
\label{Omega.Potential.modulus}
\end{equation}
The kinetic $\Omega_{\rm kin}$ and potential $\Omega_{\rm pot}$ parts
can be expressed in terms of the Green function of the neutral fermions,
which satisfies the equation
\begin{equation}
\left[-\hat I \partial_{\tau} +
\tau_{3} \left(\frac{\nabla^{2}}{2 m} + \mu \right)
+ \tau_{1} \rho \right]
{\cal G}(\tau, \mbox{\bf r}) = \delta(\tau) \delta(\mbox{\bf r}),
\label{Green.fermion.modulus}
\end{equation}
and the operator
\begin{equation}
\Sigma(\partial \theta) \equiv
\tau_{3} \left[\frac{i \partial_{\tau} \theta}{2} +
\frac{(\nabla \theta)^{2}}{8 m} \right] -
\hat{I} \left[\frac{i \nabla^{2} \theta}{4 m} +
\frac{i \nabla \theta(\tau, \mbox{\bf r}) \nabla}{2 m} \right].
                          \label{Sigma}
\end{equation}

The representation (\ref{kinetic.phase+potential}) enables
one to obtain the full set of equations necessary to find
$T_{\rm BKT}$, $\rho(T_{\rm BKT})$, and $\mu(T_{\rm BKT})$ at given
$\epsilon_{F}$ (or, for example, $\rho(T)$ and $\mu(T)$ at given $T$ and
$\epsilon_{F}$).
While the equation for $T_{\rm BKT}$ will be written using the kinetic
part (\ref{Omega.Kinetic.phase}) of the effective action, the equations
for $\rho(T_{\rm BKT})$ and $\mu(T_{\rm BKT})$ (or $\rho(T)$ and $\mu(T)$)
can be obtained using the mean field potential
(\ref{Omega.Potential.modulus}). It turns out that at a phase for which
$\rho \neq 0$, the mean-field approximation for the modulus variable
describes the system quite well. This is mainly related to the
nonperturbative character of the Hubbard--Stratonovich method, i.e., most
effects carry over for a nonzero value of $\rho$.

It is clear that the CMWH theorem does not preclude nonzero
$\langle\rho\rangle$ and, as a consequence,  an energy gap
for fermion $\chi$, since no continuous symmetry is broken when such a
gap appears. Despite  strong phase fluctuations in the two-dimensional
case, the energy gap in the spectrum of the neutral fermion $\chi$ can
still persist in the spectrum of the charged fermion $\psi$ \cite{Witten},
even well above the critical temperature.\footnote{
We note that the specific heat experiments \cite{Randeria}
demonstrated the loss of entropy that occurs at temperatures much higher than
$T_c$. This can be considered indicative of a degenerate normal state,
consistent with the existence of a nonzero order parameter $\langle\rho
\rangle$.}
We believe that the pseudogap widely discussed in high-$T_c$ cuprates
might be attributable to the energy gap of a
neutral fermion introduced in the way described above, so that the
pseudogap itself can be considered a remnant of the superconducting gap.
The condensate of neutral fermions has nothing to do with the
superconducting transition; the latter is only possible when the
superfluid density of bosons becomes large enough to stiffen the phase
$\theta(x)$. The temperature $T_\rho$ at which nonzero $\langle\rho\rangle$
develops should be identified in this approach with the pseudogap onset
temperature \cite{Randeria,Randeria.Nature}. The strategy of treating
charge and spin degrees of freedom as independent seems to be quite
useful, and at the same time a very general feature of two-dimensional
systems.

%%%%%%%%%%%%%%%%%%%%%%%%%%%%%%%%%%%%%%%%%%%%%%%%%%%%%%%%%%%%%%%%%%%%%
\section{Derivation of self-consistent  equations for $T_{\rm BKT}$,
neutral order parameter, and chemical potential}
%%%%%%%%%%%%%%%%%%%%%%%%%%%%%%%%%%%%%%%%%%%%%%%%%%%%%%%%%%%%%%%%%%%%%
\setcounter{equation}{0}

If the model under consideration is reduced to  some known model
describing the BKT phase transition, one can easily write the
equation for $T_{\rm BKT}$, which in the present approach can be identified
with the superconducting transition temperature $T_c$.  Indeed, in the
lowest orders the kinetic term (\ref{Omega.Kinetic.phase}) coincides with
the classical spin XY-model \cite{Izyumov,Minnhagen}, which has the
continuum Hamiltonian
\begin{equation}
{\cal H} = \frac{J}{2} \int d \mbox{\bf r}\,
[\nabla \theta (\mbox{\bf r})]^{2} .      \label{XY.Hamiltonian}
\end{equation}
Here $J$ is the some coefficient (in the original classical discrete XY-model
it is the stiffness of the relatively small spin rotations) and
$\theta$ is the angle (phase) of the two-component vector in the plane.

The temperature of the BKT transition is, in fact, known
for this model:
\begin{equation}
T_{\rm BKT} = \frac{\pi}{2} J.
                        \label{BKT.temperature}
\end{equation}
Despite the very simple form\footnote{An exponentially small correction is
omitted here.} of Eq. (\ref{BKT.temperature}), it was derived (see, e.g.,
Refs. \cite{Izyumov} and \cite{Minnhagen}) using the renormalization group
technique, which takes into account the non-single-valuedness of the phase
$\theta$. Thus, fluctuations of the phase are taken into account in a
higher approximation than Gaussian.  The XY-model was assumed to be
adequate for a qualitative description of the underdoped cuprates
\cite{Emery} (see also Ref. \cite{Doniach}), and the relevance of the
BKT transition to  Bose- and BCS-like superconductors
was recently discussed in Ref. \cite{Zwerger}.

To expand $\Omega _{\rm kin}$ up to $\sim (\nabla \theta)^{2}$, it is
sufficient to restrict  ourselves to terms with $n=1,2$ in the expansion
(\ref{Omega.Kinetic.phase}). The calculation is similar to
that employed in Ref. \cite{Schakel}, where only high densities $n_f$
were considered at $T=0$.  Thus, to obtain the kinetic part, one should
directly calculate the first two terms of the series
(\ref{Omega.Kinetic.phase}), which can be formally written
$\Omega_{\rm kin}^{(1)} = T \mbox{Tr} ({\cal G} \Sigma)$ and
$\Omega_{\rm kin}^{(2)} =
\frac{1}{2}T \mbox{Tr} ({\cal G} \Sigma {\cal G} \Sigma)$.  We note that
$\Sigma$ has the structure
$\Sigma = \tau_{3} O_{1} + \hat{I} O_{2}$, where $O_{1}$
and $O_{2}$ are differential operators (see (\ref{Sigma})).  One can
see, however, that the part of $\Sigma$ proportional to the unit matrix
$\hat{I}$ does not contribute to $\Omega_{\rm kin}^{(1)}$.  Hence,
\begin{equation}
\Omega_{\rm kin}^{(1)} = T \int_{0}^{\beta} d \tau \int d \mbox{\bf r}\,
\frac{T}{(2 \pi)^{2}} \sum_{n = - \infty}^{\infty}
\int d \mbox{\bf k}\, \mbox{Tr} [{\cal G}(i \omega_{n}, \mbox{\bf k})
\tau_{3}] \left( \frac{i \partial_{\tau} \theta}{2} + \frac{(\nabla
\theta)^{2}}{8 m}\right),         \label{A1} \end{equation} where
\begin{equation}
{\cal G}(i \omega_{n}, \mbox{\bf k}) = - \frac{
i \omega_{n} \hat{I} + \tau_{3} \xi(\mbox{\bf k}) - \tau_{1} \rho}
{\omega_{n}^{2} + \xi^{2}(\mbox{\bf k}) + \rho^{2}}     \label{A2}
\end{equation}
is the Green function of neutral fermions in the frequency--momentum
representation, with
$\xi(\mbox{\bf k}) = \varepsilon(\mbox{\bf k}) - \mu$ and
$\varepsilon (\mbox{\bf k}) = \mbox{\bf k}^{2}/2m$.

The summation over the Matsubara
frequencies $\omega_{n} = \pi (2n + 1) T$ and integration over
$\mbox{\bf k}$ in (\ref{A1}) can be easily performed
using the sum (\ref{B8});  one thus obtains
\begin{equation}
\Omega_{\rm kin}^{(1)} =
T \int_{0}^{\beta} d \tau \int d \mbox{\bf r}\,
n_{F}(\mu, T, \rho) \left(
\frac{i \partial_{\tau} \theta}{2} +
\frac{(\nabla \theta)^{2}}{8 m}
\right),                \label{A3}
\end{equation} where
\begin{equation}
n_{F}(\mu, T, \rho(\mu, T)) = \frac{m}{2 \pi} \left\{
\sqrt{\mu^{2} + \rho^{2}} + \mu +
2 T \ln{\left[
1 + \exp{\left(-\frac{\sqrt{\mu^{2} + \rho^{2}}}{T}\right)}
\right]}\right\}.
                 \label{A4}
\end{equation}
This has the form of a Fermi quasiparticle density
(for $\rho = 0$ the expression (\ref{A4}) is
simply the density of free fermions).

For the case $T = 0$ \cite{Thouless,Schakel}, in which real time $t$
replaces imaginary time $\tau$, one can argue from Galilean
invariance that the coefficient of $\partial_{t} \theta$
is rigorously related to the coefficient of $(\nabla \theta)^{2}$.
It therefore does not appear in $\Omega_{\rm kin}^{(2)}$.
We wish, however, to stress that these arguments cannot be used
to eliminate the term $(\nabla \theta)^{2}$
from $\Omega_{\rm kin}^{(2)}$ when $T \neq 0$, so we must calculate
it explicitly.

The $O_{1}$ term in $\Sigma$ yields
\begin{equation}
\Omega_{\rm kin}^{(2)} (O_{1})  =  \frac{T}{2}\int_{0}^{\beta} d \tau
\int d \mbox{\bf r}\,\frac{T}{(2 \pi)^{2}} \sum_{n = - \infty}^{\infty}
\int d \mbox{\bf k}\,\mbox{Tr} [{\cal G}(i \omega_{n}, \mbox{\bf k})
\tau_{3}{\cal G}(i \omega_{n}, \mbox{\bf k}) \tau_{3}]\left( \frac{i
\partial_{\tau} \theta}{2} + \frac{(\nabla \theta)^{2}}{8 m}\right)^{2}.
\label{A5}
\end{equation}
Using (\ref{B12}) to compute the sum over the Matsubara
frequencies, we find that
\begin{equation}
\Omega_{\rm kin}^{(2)} (O_{1}) = - \frac{T}{2}
\int_{0}^{\beta} d \tau \int d \mbox{\bf r}\, K(\mu, T, \rho)
\left( i \partial_{\tau} \theta +
\frac{(\nabla \theta)^{2}}{4 m}\right)^{2},
                                 \label{A6}
\end{equation}
where
\begin{equation}
K(\mu, T, \rho(\mu, T)) =
\frac{m}{8 \pi} \left(1 + \frac{\mu}{\sqrt{\mu^2 + \rho^2}}
\tanh{\frac{\sqrt{\mu^2 + \rho^2}}{2T}} \right).
                                 \label{A7}
\end{equation}

Obviously, the $O_{1}$ term does not affect the coefficient
of $(\nabla \theta)^{2}$.
Further, it is easy to make sure that the cross term involving
$O_{1}$ and $O_{2}$ in $\Omega_{\rm kin}^{(2)}$ is absent.
Finally, calculations of the $O_{2}$ contribution
to $\Omega_{\rm kin}^{(2)}$ yield\footnote{
Derivatives higher than $(\nabla \theta)^{2}$
were not computed here.}
\begin{equation}
\Omega_{\rm kin}^{(2)} (O_{2}) =  T
\int_{0}^{\beta} d \tau \int d \mbox{\bf r}\,
\frac{T}{(2 \pi)^{2}} \sum_{n = - \infty}^{\infty}
\int d \mbox{\bf k}\, \mbox{\bf k}^{2}
\mbox{Tr} [{\cal G}(i \omega_{n}, \mbox{\bf k}) \hat{I}
           {\cal G}(i \omega_{n}, \mbox{\bf k}) \hat{I}]
\frac{(\nabla \theta)^{2}}{16 m^{2}}.                 \label{A8}
\end{equation}
Thus, summing over the Matsubara frequencies (see Eq. (\ref{B13})),
one obtains
\begin{equation}
\Omega_{\rm kin}^{(2)} (O_{2}) = - \int_{0}^{\beta} d
\tau \int d \mbox{\bf r}\, \frac{1}{128 \pi^{2} m^{2}} \int d
\mbox{\bf k}\,\frac{\mbox{\bf k}^{2}}{\cosh^{2} \frac{\ds \sqrt{
\xi^{2}(\mbox{\bf k}) + \rho^{2}}}{\ds 2T}}(\nabla \theta)^{2}.                                    \label{A9}
\end{equation}
As expected, this term vanishes when $T \to 0$,
but at finite $T$ it is comparable with (\ref{A3}).

Combining (\ref{A3}), (\ref{A9}), and (\ref{A6}) we finally obtain
\begin{equation}
\Omega _{\rm kin} =
\frac{T}{2}
\int_{0}^{\beta} d \tau \int d \mbox{\bf r}
\left[n_{F}(\mu, T, \rho) i \partial_{\tau} \theta +
J(\mu, T, \rho) (\nabla \theta)^{2} +
K(\mu, T, \rho) (\partial_{\tau} \theta)^{2}
\right],
                 \label{Omega.Kinetic.phase.final}
\end{equation}
where
\begin{equation}
J(\mu, T, \rho(\mu, T)) = \frac{1}{4 m} n_{F}(\mu, T, \rho) -
\frac{T}{4 \pi} \int_{-\mu/2T}^{\infty} dx\,
\frac{x + \mu/2T}{\cosh^{2} \sqrt{x^{2} +
\frac{\ds \rho^{2}}{\ds 4 T^{2}}}}
                              \label{A10}
\end{equation}
characterizes the phase stiffness and governs the spatial
variation of the phase $\theta(\mbox{\bf r})$.
One can see that our value of the phase stiffness
$J(T=0)$ coincides with the nonrenormalized stiffness used in
Ref. \cite{Emery}.

The quantity $J(\mu,T, \rho)$ vanishes at $\rho=0$, which means that
above $T_{\rho}$ the modulus--phase variables are meaningless; to
study the model in this region one must use the old variables
$\Phi$ and $\Phi^{\ast}$. Near  $T_\rho$ one can
obtain from (\ref{A10}) in the high-density limit (see below)
\begin{equation}
J(\mu\simeq\epsilon_{F}, T \to T_{\rho}, \rho \to 0) =
\frac{7 \zeta(3)}{16 \pi^{3}} \frac{\rho^{2}}{T_{\rho}^{2}} \epsilon_{F}
\simeq 0.016 \frac{\rho^{2}}{T_{\rho}^{2}} \epsilon_{F}.
\label{J.bcs}
\end{equation}

Direct comparison of (\ref{Omega.Kinetic.phase.final}) with the Hamiltonian
of the XY-model
(\ref{XY.Hamiltonian}) makes it possible to write Eq. (\ref{BKT.temperature})
for $T_{\rm BKT}$ directly:
\begin{equation}
\frac{\pi}{2} J(\mu, T_{\rm
BKT}, \rho(\mu, T_{\rm BKT})) = T_{\rm BKT}.  \label{BKT.equation}
\end{equation}
Although mathematically this  reduces to a well-known problem,
the analogy is incomplete. Indeed, in the standard XY-model
(as well as the nonlinear $\sigma$-model) the vector (spin)
subject to ordering is assumed to be a unit vector with no dependence
on $T$.\footnote{
There is no doubt that in certain situations (for example,
very high $T$) it also can become a thermodynamic variable,
i.e., one dependent on $T$, as happens in problems of phase transitions
between ordered (magnetic) and disordered (paramagnetic) phases when
the spin itself vanishes. Specifically, for quasi-2D spin systems it is
obvious that as one proceeds from high-$T$ regions, a spin modulus first
forms in 2D clusters of finite size and only then does global 3D ordering
occur. We note, however, that this dependence
was neglected in Ref. \cite{Emery}, where the nonrenormalized phase
stiffness $J(T=0)$ was used to write Eq. (\ref{BKT.equation}).}
In our case this is definitely not the case, and a
self-consistent calculation of $T_{\rm BKT}$ as a function of $n_{f}$
requires additional equations for $\rho$ and $\mu$, which together
with (\ref{BKT.equation}) form a complete set.

Using the definition (\ref{Omega.Potential.modulus}), one
can derive the effective potential $\Omega _{\rm pot}(v, \mu, T, \rho)$
(see Appendix~A).
Then the desired missing equations are the condition
$\partial \Omega_{\rm pot}(\rho)/ \partial \rho = 0$
that the potential (\ref{B11}) be minimized, and the equality
$v^{-1} \partial \Omega_{\rm pot} /\partial \mu = -n_f$, which
fixes $n_f$.
These are, respectively
\begin{equation}
\frac{1}{V} = \int \frac{d \mbox{\bf k}}{(2 \pi)^{2}}
\frac{1}{2\sqrt{\xi^{2}(\mbox{\bf k}) + \rho^{2}}}
\tanh{\frac{\sqrt{\xi^{2}(\mbox{\bf k}) +
\rho^{2}}}{2 T}},
\label{rho}
\end{equation}
\begin{equation}
n_{F}(\mu, T, \rho) = n_f,
\label{number.rho}
\end{equation}
where $n_{F}(\mu, T, \rho)$ is defined by (\ref{A4}).

Equations (\ref{rho}) and (\ref{number.rho})
comprise a self-consistent system for determining the modulus $\rho$
of the order parameter and the chemical potential $\mu$ in the
mean-field approximation for fixed $T$ and $n_{f}$.

While Eqs. (\ref{rho}) and (\ref{number.rho}) seem to yield a reasonable
approximation at high densities $n_f$, since they include
condensed boson pairs in a nonperturbative way via nonzero $\rho$, they must
certainly be corrected in the strong coupling regime (low densities $n_f$)
to take into account the contribution of noncondensed bosons (this appears to
be important also for Eq. (\ref{BKT.equation}), which determines $T_{\rm
BKT}$).  The extent to which this alters the present results is not
completely clear. Previously, the best way to incorporate noncondensed pairs
seems to have been the self-consistent $T$-matrix approximation
\cite{Serene,Tchernyshyov,Levin,Nazarenko}, which allows one to account for
the feedback of pairs on the self-energy of fermions.  However, the
$T$-matrix approach, at least in its standard form
\cite{Serene,Tchernyshyov,Levin,Nazarenko},
fails to describe the BKT phase transition, for which one must consider the
equation for the vertex. On the other hand, in our approach the BKT phase
transition is realized by the condition (\ref{BKT.temperature}), while an
analog of the $T$-matrix approximation in terms of propagators of the
$\rho$-particle and the neutral fermion $\chi$ has yet to be elaborated.

The energy of two-particle bound states in a vacuum
\begin{equation}
\varepsilon_{b} = -2W \exp
\left(-\frac{4 \pi}{m V} \right)
\label{bound.energy}
\end{equation}
(see Refs. \cite{Randeria.book,GGL} and \cite{Miyake}) is more convenient to
use than the four-fermion constant $V$ (here $W$ is the conduction
bandwidth).  For example, one can easily take the limits $W \to \infty$ and
$V \to 0$ in Eq. (\ref{rho}), which after this renormalization
becomes
\begin{equation}
\ln{\frac{|\varepsilon_{b}|}{\sqrt{\mu^{2} +
\rho^{2}} - \mu }} = 2 \int_{-\mu/T}^{\infty} d u\, \frac{1}{\ds \sqrt{u^2 +
\left( \frac{\rho}{ T} \right)^2} \left[\exp{\sqrt{u^2 + \left(
\frac{\rho}{T} \right)^2}} + 1 \right]} .
\label{rho.bound}
\end{equation}
Thus, in practice, we solve Eqs. (\ref{BKT.equation}),
(\ref{number.rho}), and (\ref{rho.bound}) numerically to study $T_{\rm BKT}$
as function of $n_{f}$ (or equivalently, of the Fermi energy $\epsilon_{F} =
\pi n_{f}/m$, as it should be for 2D metals with the simplest quadratic
dispersion law).

It is easy to show that at $T = 0$, the system (\ref{number.rho}),
(\ref{rho.bound})  transforms into a previously studied system
(see Ref. \cite{Randeria.book} and references therein). Its
solution is $\rho = \sqrt{2 |\varepsilon_{b}| \epsilon_{F}}$ and $\mu = -
|\varepsilon_{b}|/2 + \epsilon_{F}$. This will be useful in studying the
concentration dependencies of $2 \Delta/T_{\rm BKT}$ and $2 \Delta/T_{\rho}$,
where $\Delta$ is the zero-temperature gap in the quasiparticle excitation
spectrum. It should be borne in mind that in the local pair regime
($\mu < 0$), the gap $\Delta$ equals $\sqrt{\mu^{2} + \rho^{2}}$ rather than
$\rho$ (as in the case $\mu > 0$) \cite{Randeria.book}.

Setting $\rho = 0$ in Eqs. (\ref{rho}) and
(\ref{number.rho}), we obtain (in the same approximation) the
equations for the critical temperature $T_{\rho}$ and the corresponding
value of $\mu$:
\begin{equation}
\ln{\frac{|\varepsilon_{b}|}{T_{\rho}}\frac{\gamma}{\pi}} =
- \int_{0}^{\mu/2T_{\rho}} d u\, \frac{\tanh{u}}{u}
\qquad (\gamma = 1.781),
\label{temperature.rho}
\end{equation}
\begin{equation}
T_{\rho} \ln{\left[1 +
\exp{ \left( \frac{\mu}{T_{\rho}} \right)}\right]} = \epsilon_{F}.
\label{number.temperature.rho}
\end{equation}
\noindent
Note that these equations coincide with the system that determines
the mean-field temperature $T_{c}^{(2D) MF}\,(=T_\rho)$ and
$\mu(T_{c}^{(2D) MF})$  \cite{GGL}, evidently as a result of the
mean-field approximation for the variable $\rho$ used here.  There is,
however, an important difference between the temperatures $T_c^{2D}$ and
$T_\rho$.  Specifically, if one takes fluctuations into account, $T_{c}^{2D}$
goes to zero, while the value of $T_{\rho}$ remains finite. The crucial point
is that the perturbation theory in the variables $\rho$ and $\theta$ does not
contain any infrared singularities \cite{Witten,Ichinose}, in contrast to the
perturbation theory in $\Phi,\Phi^{\ast}$; thus the fluctuations do not
reduce $T_\rho$ to zero.  This is why the temperature $T_{\rho}$ has its own
physical meaning: incoherent (local or Cooper) pairs begin to form (at least
at high enough $n_f$ (see Sec.~V)) just below $T_{\rho}$.  At higher
temperatures, only these pair fluctuations exist; their influence was
studied in Ref. \cite{LSh.FNT.1997}.

%%%%%%%%%%%%%%%%%%%%%%%%%%%%%%%%%%%%%%%%%%%%%%%%%%%%%%%%%%%%%%%%%%%%%
\section{Numerical results}
%%%%%%%%%%%%%%%%%%%%%%%%%%%%%%%%%%%%%%%%%%%%%%%%%%%%%%%%%%%%%%%%%%%%%
\setcounter{equation}{0}

A numerical investigation of the systems (\ref{BKT.equation}),
(\ref{number.rho}), (\ref{rho.bound}), and (\ref{temperature.rho}),
(\ref{number.temperature.rho}) yields the following
results, which are displayed graphically as the phase diagram of the
system.

{\rm a)} For low carrier densities, the pseudogap phase area
(see Fig.~1) is comparable with the BKT area.
For high carrier densities ($\epsilon_{F} \rtwid 10^3|\varepsilon_{b}|$),
one easily finds that the pseudogap region shrinks asymptotically
as
\begin{equation}
\frac{T_{\rho} - T_{\rm BKT}}{T_{\rho}} \simeq
\frac{4 T_{\rho}}{\epsilon_{F}}.
             \label{asymp}
\end{equation}
This behavior  qualitatively restores the BCS limit observed
in overdoped samples.

{\rm b)} For $\epsilon_{F} \leq (10\,-\,15) |\varepsilon_{b}|$, the
function $T_{\rm BKT}(\epsilon_{F})$ is linear, as also
confirmed by the analytic solution of the system
(\ref{BKT.equation}), (\ref{number.rho}), and (\ref{rho.bound}), which
yields $T_{\rm BKT} = \epsilon_{F}/8$.  Remarkably, such behavior of
$T_{c}(\epsilon_{F})$ is observed for all families of HTSC
cuprates in their underdoped region \cite{Uemura,Emery}, though with
a smaller coefficient of proportionality ($0.01\,-\, 0.1$). This
indicates the importance of including a contribution due to
noncondensed pairs in Eq. (\ref{BKT.equation}), which defines $T_{\rm
BKT}$.

It has been shown that for optimal doping, the dimensionless ratio
$\epsilon_{F}/|\varepsilon_{b}| \sim 3 \cdot10^{2}\,-\, 10^{3}$
\cite{Carter}.  Thus it is quite natural to suppose that in the underdoped
region one has $\epsilon_{F}/|\varepsilon_{b}| \sim  10\,-\, 10^{2}$, where
we find linear behavior.

We note that in this limit, the temperature $T_{c}$ of formation of
a homogeneous order parameter for the quasi-2D model
\cite{Uemura,GLSh.PhysicaC} can easily be written in the form
\begin{equation}
T_{c} \approx \frac{4 T_{\rm BKT}}
{\ln (\epsilon_{F} |\varepsilon_{b}|/4 t_{\mid \mid}^{2})},
                     \label{temperature.quasi}
\end{equation}
where $t_{\mid \mid}$ is the interplane hopping (coherent tunneling)
constant.  This shows that when $T_{c} < T_{\rm BKT}$, the weak
three-dimensionalization can preserve (in any case, at low $n_{f}$) the
regions of the pseudogap and BKT phases, which, for example, happens in the
relativistic quasi-2D model \cite{Ichinose}. At the same time, as the
three-dimensionalization parameter $t_{\mid \mid}$ increases, when $T_{c} >
T_{\rm BKT}$ the BKT phase can vanish, provided, however, that the anomalous
phase region and both temperatures $T_{\rho}$ and $T_{c} \simeq n_f/m$ are
preserved.

{\rm c)} Figure~2 shows the values of $n_{f}$ for which $\mu$ differs
substantially from $\epsilon_{F}$, or in other words, the Landau
Fermi-liquid theory becomes inapplicable to metals ( also called bad ones)
with low or intermediate carrier density. As expected, the kink $\mu$ at
$T = T_{\rho}$,  which has been observed experimentally \cite{Marel} and
interpreted for the 1-2-3 cuprates \cite{Dotsenko}, becomes
less and less pronounced as $\epsilon_{F}$ increases. But in the present
case it is interesting that in the hydrodynamic approximation employed
here, it happens at the normal--pseudogap phase boundary or before
superconductivity really appears. It would therefore be of great interest to
perform experiments that might reveal the temperature dependence $\mu(T)$,
especially for strongly anisotropic and relatively weakly doped cuprates.

{\rm d)} It follows from curve 3 in Fig.~2 that the crossover
(sign change in $\mu$) from local to Cooper pairs is possible not
only as $\epsilon_{F}$ increases, which is more or less obvious, but
also (for some $n_{f}$) as $T$ increases.

{\rm e)} Finally the calculations showed (see Fig.~3) that the ratio
$2 \Delta/T_{\rm BKT}$ is greater than 4.7 in the region under study.
The value $2 \Delta/T_{\rho} (= 2\Delta/T_{c}^{MF})$ is, however,
somewhat lower and reaches the BCS theory limit of 3.52 only for
$\epsilon_{F} \gg |\varepsilon_{b}|$. It is interesting that this
concentration behavior is consistent with numerous measurements of
this ratio in HTSC \cite{Devereaux,Kendziora}. Note that the divergence
of $2 \Delta/T_{\rm BKT}$ and $2 \Delta/T_{\rho}$ at $\epsilon_{F} \to 0$
is directly related to the definition of $\Delta$ at $\mu < 0$.

%%%%%%%%%%%%%%%%%%%%%%%%%%%%%%%%%%%%%%%%%%%%%%%%%%%%%%%%%%%%%%%%%%%%%
\section{Pairing temperature $T_{P}$ versus carrier density}
%%%%%%%%%%%%%%%%%%%%%%%%%%%%%%%%%%%%%%%%%%%%%%%%%%%%%%%%%%%%%%%%%%%%%
\setcounter{equation}{0}

There is no disagreement concerning the asymptotic behavior of
$T_{\rm BKT}$ (or $T_{c}$) $\sim \epsilon_{F}$ in the region of
low carrier densities. In contrast, the behavior of the temperature
$T_{\rho}$, below which pairs are formed, cannot be considered to be
generally accepted. For example, in Refs. \cite{Uemura} and \cite{Emery},
based on qualitative arguments, this temperature is taken to be
 the temperature $T_{P}$ of local uncorrelated pairing,
which in contrast to $T_{\rho}$ increases with decreasing $n_{f}$.\footnote
{In fact, in Refs. \cite{Uemura} and \cite{Emery} (see also
Ref. \cite{Haussmann}) this temperature was plotted as an increasing function
of coupling constant $V$, which for 3D systems corresponds, to some extent,
to the carrier density decreasing.  In 2D systems, however, where, as is
well known, two-particle bound states are formed without any threshold,
similar conclusions about the behavior of $T_P(n_f)$ are questionable, and
must be checked independently.}
Randeria (see Ref. \cite{Randeria.book} and references therein), to define
the pairing temperature $T_{P}$, uses the system
of equations for the mean-field transition temperature and the corresponding
chemical potential, which is essentially identical to the system
(\ref{temperature.rho}), (\ref{number.temperature.rho}).  Thus his $T_{P}
\to 0$ as $n_{f} \to 0$.

It is also well known \cite{Randeria.book,Haussmann,Varma} that in the
low-density limit, it is vital to include quantum fluctuations, at least
in the number equation \cite{Nozieres}, in the calculation of
the critical temperature at which long-range order forms in 3D. In 2D
these fluctuations in fact reduce the critical temperature to zero
\cite{Tokumitu}. Certainly quantum fluctuations are also important in
the calculation of $T_\rho$ in the limit $n_f \to 0$ and, in particular,
in the number equation.  However, as already stressed in Sec.~III, these
corrections are quite different from what we obtain using the variables
$\Phi,\Phi^{\ast}$, since perturbation theory in the variables $\rho$ and
$\theta$ does not contain any infrared singularities \cite{Witten,Ichinose},
and the fluctuations do not yield $T_\rho\equiv0$.  In fact, even including
quantum fluctuations, $T_{\rho}$ must exceed $T_{\rm BKT}$ ($\rho(T_{\rm
BKT}) \ne 0$), so that the pseudogap phase is always present.

In our opinion, the temperature $T_{\rho}$ has its own physical
interpretation: this is the temperature of a smooth transition to the
state in which the neutral order parameter $\rho \neq 0$, and below which
one can observe pseudogap manifestations. There is also a very interesting
and important question about the character of the transition. Certainly in
the simplest Landau theory one appears to have a second-order phase
transition, since $\rho$ takes a nonzero value only below $T_{\rho}$
\cite{MacKenzie}. However this kind of transition is only possible for
neutral fermions.  Fluctuations of the $\theta$-phase will  transform the
pole in the Green function of the neutral fermions into a branch cut in the
Green function for charged particles in the BKT phase. Indeed, the CMWH
theorem concerning the absence of spontaneous breaking of a continuous
symmetry means that symmetry-violating Green functions must vanish. However,
it says nothing about the gap in the spectrum of excitations, as is
sometimes incorrectly stated.

The correct explanation is that if the symmetry is unbroken, {\it and} the
fermion excitation appears as a pole in the $\psi$ two-point function, then
the fermion must be gapless.  If the fermion does not have the same quantum
numbers as $\psi$ (like our fermion $\chi$) and so does not appear in the
$\psi$ two-point function as a one-particle state, then the symmetry does not
tell  whether the fermion ($\chi$) will be gapless or not.

This very general argument \cite{Witten} suggests the
following plausible scenario. At low temperatures ($T<T_{\rm BKT}$),
$\chi,\rho,$ and $\theta$ should be treated as physical quasiparticles
($\chi,\rho$ having a gap and $\theta$ being a gapless excitation), while a
straightforward computation of the $\psi$ two-point function
\cite{Witten} reveals its branch-cut structure.

On the other hand, at temperatures above $T_{\rm BKT}$, we should consider
$\psi$ and $\Phi$ true quasiparticles, since $T_{\rm BKT}$ is a phase
transition point and the spectrum of physical excitations changes precisely
at this point.  The $\psi$ two-point function at $T>T_{\rm BKT}$ should be
studied separately due to the presense of vortices which change the form of
the correlator $\langle \exp[i \theta(x)] \exp[i \theta(0)] \rangle$ above
$T_{\rm BKT}$. In this temperature region the $\psi$ two-point function
loses its branch-cut structure; instead, it acquires the form suggested in
Refs.\cite{Tchernyshyov} and \cite{Levin} with a pseudogap originating from
the superconducting gap below $T_{\rm BKT}$, which preserves ``BCS-like''
structure as well as the diagonal component of the single-particle Green
function. In this picture the Fermi-liquid description breaks down, evidently
below $T_\rho$, due to the formation of nonzero $\rho$.

We note, however, that the decisive confirmation of this picture demands
further detailed study probably based on a different approach, for example
the self-consistent $T$-matrix (see Ref. \cite{Tchernyshyov} and
references therein), which enables one to directly obtain the full
fermion Green function.

To define the temperature $T_{P}$ properly, one should study the spectrum of
bound states either by solving the Bethe--Salpeter equation \cite{GGL} or by
analyzing the corresponding Green functions as we do here.  It turns out
that there is no difference between $T_{P}$ and $T_{\rho}$ in the Cooper pair
regime ($\mu > 0$), while in the local pair region ($\mu < 0$) these
temperatures exhibit different behavior.

Indeed, let us study the spectrum of bound states in both the normal
($\rho = 0$) and pseudogap ($\rho \neq 0$) phases. We are especially
interested in determining the conditions under which  real bound states
(with zero total momentum $\mbox{\bf K} = 0$) become unstable.  For this
purpose one can look at the propagator of the $\rho$-particle in the
pseudogap phase:
\begin{equation}
\Gamma^{-1}(\tau, \mbox{\bf r}) =  \left. \frac{1}{2}
\frac{\beta \delta^{2} \Omega(v, \mu,T, \rho(\tau, \mbox{\bf r}),
\partial \theta(\tau, \mbox{\bf r}))}
{\delta \rho(\tau, \mbox{\bf r}) \delta \rho(0, 0)}
\right|_{\rho = \rho_{\rm min} = const},
\label{Gamma.rho}
\end{equation}
where $\rho_{\rm min}$ is defined by the minimum condition
(\ref{rho}) (or (\ref{rho.bound})) of the potential part (\ref{B11}) of the
effective action (\ref{Effective.Action.2D}).
In the momentum representation, the spectrum of bound states is
usually determined by the condition
\begin{equation}
\Gamma^{-1}_{R} (\omega, \mbox{\bf K}) = 0,
\label{spectrum}
\end{equation}
where $\Gamma_{R} (\omega, \mbox{\bf K})$ is the retarded Green
function obtained directly from the temperature Green function
$\Gamma (i\Omega_{n}, \mbox{\bf K})$ using the analytic continuation
$i\Omega_{n} \to \omega + i0$.
Recall that such analytic continuation must be performed after
evaluating the sum over the Matsubara frequencies. In the case of
vanishing total momentum $\mbox{\bf K} = 0$, one arrives at the energy
spectrum equation
\begin{equation} \Gamma^{-1}_{R} (\omega, 0) = \frac{1}{V} + 2 \int
\frac{d \mbox{\bf k}}{(2 \pi)^{2}} \frac{\xi^{2}(\mbox{\bf
k})}{\sqrt{\xi^{2}(\mbox{\bf k}) + \rho^{2}}} \frac{\tanh \sqrt{
\xi^{2}(\mbox{\bf k}) + \rho^{2}}/2T} {\omega^{2} - 4 [\xi^{2}
(\mbox{\bf k}) + \rho^{2}]} = 0.
\label{spectrum.rho}
\end{equation}
From the explicit
expression (\ref{spectrum.rho}) for $\Gamma_{R} (\omega, 0)$,  this function
obviously has a branch cut at frequencies
\begin{equation} |\omega| \geq 2\,
\mbox{min} \sqrt{ \xi^{2}(\mbox{\bf k}) + \rho^{2}} =
\left\{\begin{array}{cc} 2 \rho,  \qquad \quad & \mu \geq 0 \\ 2
\sqrt{\mu^{2} + \rho^{2}}, & \mu < 0.  \end{array} \right.
\label{cut}
\end{equation}
Thus, bound states can exist below this cut.

Real bound states decay into two-fermion states
when the energy of the former reaches the branch point
$2\,\mbox{min} \sqrt{ \xi^{2}(\mbox{\bf k}) + \rho^{2}}$. Since
$\Gamma_{R}^{-1}$ is a monotonically decreasing function of
$\omega^{2}$, it has the unique solution
$|\varepsilon_{b}(T)| = 2 \rho(T)$, at which Eq. (\ref{spectrum.rho})
coincides exactly with the mean-field equation (\ref{rho}) for $\rho(T)$.
It also becomes clear that for $\mu < 0$ we have real bound states with
energy $\varepsilon_{b}(T)$ below the two-particle scattering continuum at
$\omega = 2\sqrt{\mu^{2} + \rho^{2}}$, while at $\mu \geq 0$ there
are no stable bound states. The line $\mu(T, \epsilon_{F}) =0$
in the $T$--$\epsilon_{F}$ plane at $\rho \neq 0$ separates the
negative $\mu$ region where local pairs exist from that
in which only Cooper pairs exist (positive $\mu$).
This line (see Fig.~4) begins at the point
$T = (e^{\gamma}/\pi) |\varepsilon_{b}| \approx 0.6 |\varepsilon_{b}|$,
$\epsilon_{F} \approx 0.39 |\varepsilon_{b}|$ and ends at $T=0$,
$\epsilon_{F} = |\varepsilon_{b}|/2$. (The latter follows directly
from the solution at $T=0$, $\mu = -|\varepsilon_{b}|/2 + \epsilon_{F}$
\cite{Randeria.book,GGL}.)

To find a similar line in the normal phase with $\rho=0$, we consider the
corresponding equation for the bound states. The propagator of
these states (in imaginary time formalism) is defined to be
\begin{equation}
\Gamma^{-1}(\tau, \mbox{\bf r}) =  \left.
\frac{\beta \delta^{2} \Omega(v, \mu,T, \Phi(\tau, \mbox{\bf r}),
\Phi^{\ast}(\tau, \mbox{\bf r}))}
{\delta \Phi^{\ast}(\tau, \mbox{\bf r}) \delta \Phi(0, 0)}
\right|_{\Phi = \Phi^{\ast} = 0}.
\label{Gamma.Phi.def}
\end{equation}
(In the normal phase, where $\rho =0$, we must again use  the initial
auxiliary fields $\Phi$ and $\Phi^{\ast}$ (see Secs.\ II. and III).) Then in
the momentum representation (after summing  over the Matsubara frequencies)
we have
\begin{eqnarray}
\Gamma^{-1}(i\Omega_{n}, \mbox{\bf K}) =
\frac{1}{V} - \frac{1}{2} \int \frac{d \mbox{\bf k}}{(2 \pi)^{2}}
\frac{\tanh \xi_{+}(\mbox{\bf k},\mbox{\bf K})/2T +
      \tanh \xi_{-}(\mbox{\bf k},\mbox{\bf K})/2T}
{\xi_{+}(\mbox{\bf k},\mbox{\bf K}) + \xi_{-}(\mbox{\bf k},\mbox{\bf K})
- i \Omega_{n}},
\nonumber                                    \\
\xi_{\pm}(\mbox{\bf k},\mbox{\bf K}) \equiv \frac{1}{2m}
\left(\mbox{\bf k} \pm \frac{\mbox{\bf K}}{2} \right)^2 - \mu,
                      \label{Gamma.Phi}
\end{eqnarray}
where $\mbox{\bf k}$ is the relative momentum of the pair. The
spectrum of bound states is given again by Eq. (\ref{spectrum}).
Using the energy $\varepsilon_{b}$ (see Eq. (\ref{bound.energy})) of
the bound state at $T = 0$, for $\mbox{\bf K} = 0$ we obtain the
following equation for the energies of these states in the normal phase:
\begin{equation}
\int_{0}^{\infty} dx \left[\frac{1}{x + |\varepsilon_{b}|/2} -
\frac{\tanh(x - \mu)/2T}{x - \mu - \omega/2}\right] = 0.
\label{spectrum.Phi}
\end{equation}
Such states can exist provided
$-2 \mu - |\varepsilon_{b}| < \omega < - 2 \mu$. The left-hand side
of Eq. (\ref{spectrum.Phi}) is positive at
$\omega = -2\mu - |\varepsilon_{b}|$ and tends to $+ \infty$
($\mu > 0$) or $- \infty$ ($\mu < 0$)  when $\omega \to - 2 \mu$.
This equation always has a solution at $\mu < 0$, so
 bound states with zero total momentum exist for negative $\mu$.

For $\mu > 0$, analytic analysis becomes more complicated, and requires
numerical study.  One can easily find from (\ref{spectrum.Phi}) that at $T
=0$, stable bound states exist up to $\mu < |\varepsilon_{b}|/8$. In fact,
numerical study for $T \geq T_{\rho}$ shows that the trajectory $\mu (T,
\epsilon_{F}) = 0$ (or $T = \epsilon_{F}/ \ln2$, see
(\ref{number.temperature.rho})) approximately divides the normal phase into
two qualitatively different regions -- with ($\mu < 0$) and without ($\mu >
0$) stable (long-lived) pairs. This also holds for
other phases, which enables one to draw the whole line $\mu(T,
\epsilon_{F})=0$ (Fig.~4).

Knowing the two-particle binding energy, it is natural to define
pairing temperature $T_{P}$ as
$T_{P} \approx |\varepsilon_{b}(T_{P}, \mu(T_{P},\epsilon_{F}))|$.
This equation  can be easily analyzed in the region
$\epsilon_{F} \ll |\varepsilon_{b}|$, for which we directly obtain
$T_{P} \approx |\varepsilon_{b}|$, which clearly coincides with the standard
estimate \cite{Uemura,Micnas}. This means in turn that the curve
$T_{P}(\epsilon_{F})$ starting at $T_{P}(0) \approx |\varepsilon_{b}|$
will be reduced, up to the point
$T_{P}(0.39 \epsilon_{F}) \approx 0.6 |\varepsilon_{b}|$, which lies
on the line $T_{\rho}(\epsilon_{F})$ (see Fig.~4). It is important
that this line is not the phase transition curve; it merely divides
the fermion system diagram into temperature regions with a prevailing mean
number of local pairs ($T \ltwid T_{P}$) or unbound carriers ($T \rtwid
T_{P}$).  This is the region of the abnormal normal phase where one has
preformed boson pairs. It is widely accepted, however, that this case is only
of theoretical interest, since there is no Fermi surface ($\mu<0$) in the
phase. The phase area or the difference $T_P(\epsilon_F)-
T_{\rho}(\epsilon_F)$ is an increasing function as $\epsilon_{F} \to 0$,
which corresponds to the behavior usually assumed \cite{Uemura,Emery}.

When $\mu>0$ there are no stable bound states
($\varepsilon_{b}(T) = 2 \rho(T) =0$) for the normal phase, where they
are short-lived. Formally, using $\rho(T)=0$ in Eq. (\ref{spectrum.rho}), we
immediately obtain (\ref{temperature.rho}) or, in other words, here $T_{P} =
T_{\rho}$. Such a conclusion is in accordance with the generally accepted
definition of $T_{P}$ in the BCS case \cite{Micnas}.

Thus the phase diagram of a 2D metal above $T_{c}$ acquires the form
shown in Fig.~4. It is interesting that if the line
$T_{P}(\epsilon_{F})$ cannot be defined exactly, the temperature
$T_{\rho}(\epsilon_{F})$ is the line below which pairs reveal some signs
of collective behavior. Moreover,
at $T < T_{\rho}$ one can speak of a real pseudogap in the one-particle
spectrum, while in the region $T_{\rho} < T < T_{P}$ only strongly
developed pair fluctuations (some number of pairs) exist, though they
probably suffice to reduce the  spectral quasi-particle
weight, and to produce other observed manifestations that mask pseudogap
(spin gap; see Ref. \cite{LSh.FNT.1997}) formation.

%%%%%%%%%%%%%%%%%%%%%%%%%%%%%%%%%%%%%%%%%%%%%%%%%%%%%%%%%%%%%%%%%%%%%
\section{Paramagnetic susceptibility of the system}
%%%%%%%%%%%%%%%%%%%%%%%%%%%%%%%%%%%%%%%%%%%%%%%%%%%%%%%%%%%%%%%%%%%%%
\setcounter{equation}{0}

It would be very interesting to study how a nonzero value of
the neutral order parameter affects the observable properties of the
2D system. Does this really resemble the gap opening in the traditional
superconductors, except that it happens in the normal phase? Or,
in other words, does the pseudogap open?

We shall demonstrate this phenomenon, taking the paramagnetic susceptibility
of the system as the simplest case in point.
To study the system in the magnetic field $\mbox{\bf H}$
one must add the paramagnetic term
\begin{equation}
{\cal H}_{PM} = - \mu_{B} H \left[
\psi^{\dagger}_{\uparrow}(\mbox{\bf r}) \psi_{\uparrow}(\mbox{\bf r}) -
\psi^{\dagger}_{\downarrow}(\mbox{\bf r}) \psi_{\downarrow}(\mbox{\bf r})
\right]
\label{Hamiltonian.magnetic}
\end{equation}
to the Hamiltonian (\ref{Hamilton}) where $\mu_{B} = e \hbar/2mc$ is the
Bohr magneton.
Note that, using the isotropy in the problem,
we chose the direction of field $\mbox{\bf H}$ to be perpendicular to
the plane containing the vectors $\mbox{\bf r}$.

Adding the corresponding term to Eq. (\ref{Green.fermion.modulus})
for the neutral fermion Green function, it is easy to show that in the
momentum representation (compare with (\ref{A2}))
\begin{equation} {\cal G}(i
\omega_{n}, \mbox{\bf k}, H)  = \frac{(i \omega_{n} + \mu_{B} H) \hat I +
\tau_{3} \xi(\mbox{\bf k}) - \tau_{1} \rho}
{(i \omega_{n} + \mu_{B} H)^{2} - \xi^{2} (\mbox{\bf k}) - \rho^{2}}.
\label{Green.momentum.magnetic}
\end{equation}

The static paramagnetic susceptibility can be expressed
in terms of the magnetization,
\begin{equation}
\chi(\mu, T, \rho) =  \left. \frac{\partial M(\mu, T, \rho, H)}
{\partial H} \right|_{H = 0},
\label{susceptibility.definition}
\end{equation}
which in the mean-field approximation can be derived from the
effective potential:
\begin{equation}
M(\mu, T, \rho, H) = - \frac{1}{v}
\frac{\partial \Omega_{\rm pot}(v, \mu, T, \rho, H)}{\partial H}.
\label{magnetization.definition}
\end{equation}
Thus from (\ref{magnetization.definition}) one obtains
\begin{equation}
M(\mu, T, \rho, H) = \mu_{B} T \sum_{n = - \infty}^{\infty}
\int \frac{d \mbox{\bf k}}{(2 \pi)^{2}}\, \mbox{Tr}
[{\cal G} (i \omega_{n}, \mbox{\bf k}, H) \hat I].
\label{magnetization}
\end{equation}
Then using the definition (\ref{susceptibility.definition}) one
arrives at
\begin{equation}
\chi(\mu, T, \rho) = \mu_{B}^{2}
\int \frac{d \mbox{\bf k}}{(2\pi)^{2}}\,
2 T \sum _{n = - \infty}^{\infty}
\frac{\xi^{2}(\mbox{\bf k}) + \rho^{2} - \omega_{n}^{2}}
{[\omega_{n}^{2} + \xi^{2}(\mbox{\bf k}) + \rho^{2}]^{2}}.
\label{susceptibility.sum}
\end{equation}
The sum in (\ref{susceptibility.sum}) can easily be calculated with the help
of Eq. (\ref{B12}); thus, we obtain the final result
\begin{equation}
\chi(\mu, T, \rho) = \chi_{\rm Pauli} \frac{1}{2}
\int_{-\mu/2T}^{\infty}
\frac{dx}{\cosh^{2} \sqrt{x^{2} + \frac{\ds \rho^{2}}{\ds 4T^{2}}}},
\label{susceptibility.final}
\end{equation}
where $\chi_{Pauli} \equiv \mu_{B}^{2} m/\pi$ is the Pauli
paramagnetic susceptibility for the 2D system.

To study $\chi$ as a function of $T$ and $n_{f}$
(or $\epsilon_{F}$), Eq. (\ref{susceptibility.final}) should
be used together with Eqs. (\ref{number.rho}) and (\ref{rho.bound}).

For the case of the normal phase ($\rho = 0$) one can
investigate the system analytically. Thus (\ref{susceptibility.final})
takes the form
\begin{equation}
\chi(\mu, T, \rho =0) = \chi_{\rm Pauli} \frac{1}{1 + \exp(-\mu/T)},
\label{susceptibility.NP.mu}
\end{equation}
where $\mu$ is determined by (\ref{number.temperature.rho}).
This system has the solution
\begin{equation}
\chi(\epsilon_{F}, T, \rho = 0) = \chi_{\rm Pauli}
[1 - \exp(-\epsilon_{F}/T)],
\label{susceptibility.NP}
\end{equation}
which is identical to a solution known from the literature
\cite{Kvasnikov}.

The results of a numerical study of the system
(\ref{susceptibility.final}), (\ref{rho.bound}), and (\ref{number.rho})
are presented in Fig.~5. One can see that the kink in $\chi$
occurs at $T = T_{\rho}$ as in the dependence of $\mu$ on $T$.
Below $T_{\rho}$ the value of $\chi(T)$ decreases, although the system is
still normal. This can be interpreted as a spin-gap (or pseudogap) opening.
The size of the pseudogap region depends strongly on the doping
($\epsilon_{F}/ |\varepsilon_{b}|$), as observed for real
HTSC \cite{Randeria,Pines.review,Randeria.Nature}. For small values of
$\epsilon_{F}/ |\varepsilon_{b}|$ this region is very large
($T_{\rho} > 6T_{\rm BKT}$), while for large
$\epsilon_{F}/ |\varepsilon_{b}| \sim 5\,-\,30$ it is slightly
larger than the region corresponding to the BKT phase.

%%%%%%%%%%%%%%%%%%%%%%%%%%%%%%%%%%%%%%%%%%%%%%%%%%%%%%%%%%%%%%%%%%%%%
\section{Conclusion}
%%%%%%%%%%%%%%%%%%%%%%%%%%%%%%%%%%%%%%%%%%%%%%%%%%%%%%%%%%%%%%%%%%%%%

To summarize, we have discussed the crossover in the
superconducting transition between BCS- and Bose-like behavior for
the simplest 2D model, with s-wave nonretarded attractive interaction.

While there is still no generally accepted microscopic theory of HTSC
compounds and their basic features (including the pairing mechanism), it
seems that this approach, although in a sense phenomenological, is
of great interest since it is able to cover the whole range of carrier
concentrations (and thus the whole range of coupling constants) and
temperatures. As we tried to demonstrate, it enables one to propose both
a reasonable interpretation for the  observed phenomena caused by doping
and to describe new phenomena---for example, pseudogap phase formation
as a new thermodynamically equilibrium normal state of low-dimensional
conducting electronic systems.

Evidently there are a number of important open questions.
They may be divided into two classes: the first concerns the problem
of a better and more complete treatment of the models themselves.
The second class relates to the  extent to which this
model is applicable to HTSC compounds, and what the necessary
ingredients are for a more realistic description.

Regarding the microscopic Hamiltonian as a given model,
our treatment is obviously still incomplete. In particular,
there exists an unconfirmed numerical result \cite{Deisz} based on a fully
self-consistent determination of a phase transition to a superconducting
state in a conserving approximation, which states that the superconducting
transition is neither the simple mean-field transition nor the BKT
transition. (See, however, the discussion preceding Eq. (\ref{bound.energy}).)
Besides, it would be very interesting to obtain the
spectrum of both the anomalous normal and pseudogap phases. It is
important also to take into consideration the effects of noncondensed bosons,
which might help to obtain a smaller slope in the dependence of $T_{\rm
BKT}$ on $\epsilon_{F}$.

As for the extent to which the models considered are really applicable to
HTSC, most of the complexity of these systems is obviously neglected here.
For example, we did not take into account the indirect nature of attraction
between the fermions, {\it d}-wave pairing, inter-layer tunneling, etc.
Nevertheless, one may hope that the present simple model
can explain the essential features of pseudogap formation.

\section*{Acknowledgments}
We thank Drs. E.~V.~Gorbar, I.~A.~Shovkovy, O.~Tchernyshyov, and
V.~M.~Turkowski for fruitful discussions, which helped to clearify some
deep questions about low-dimensional phase transitions. We especially
thank Prof. R.~M.~Quick for many thoughtful comments on an earlier version
of this manuscript.  One of us (S. G. S) is grateful to the members
of the Department of Physics of the University of Pretoria, especially Prof.
R.~M.~Quick and Dr.  N.~J.~Davidson, for very useful points and hospitality.
S.G.S also acknowledges the financial support of
the Foundation for Research Development, Pretoria.

\renewcommand{\theequation}{\Alph{section}.\arabic{equation}}
\appendix

\section{calculation of the effective potential}
\setcounter{equation}{0}
Here we sketch the derivation of the effective potential.
To obtain it one must write Eq. (\ref{Omega.Potential.modulus})
in the momentum representation:
\begin{eqnarray}
&& \Omega_{\rm pot}(v, \mu, T, \rho) = v \left\{\frac{\rho^{2}}{V}
- T \sum_{n = -\infty}^{+\infty}\int \frac{d {\mbox{\bf k}}}{(2
\pi)^{2}}\,\mbox{Tr} [\ln {\cal G}^{-1} (i \omega_{n}, \mbox{\bf k})
e^{i \delta \omega_{n} \tau_{3}}]\right.\nonumber\\
&& \qquad + \left. T \sum_{n = -\infty}^{+\infty}
\int \frac{d {\mbox{\bf k}}}{(2 \pi)^{2}}\,
\mbox{Tr} [\ln G_{0}^{-1} (i \omega_{n}, \mbox{\bf k})
e^{i \delta \omega_{n} \tau_{3}}] \right\}, \quad
\delta \to +0,
\label{B1}
\end{eqnarray}
where
\begin{equation}
{\cal G}^{-1} (i \omega_{n}, \mbox{\bf k}) =
i \omega_{n} \hat I - \tau_{3} \xi(\mbox{\bf k}) +
\tau_{1} \rho,\quad
 G_{0}^{-1} (i \omega_{n}, \mbox{\bf k}) =
\left. {\cal G}^{-1} (i \omega_{n}, \mbox{\bf k})
\right|_{\rho = \mu =0}
\label{B2}
\end{equation}
are the inverse Green functions.
The exponential factor  $e^{i \delta \omega_{n} \tau_{3}}$ is added
to (\ref{B1}) to provide the correct regularization which is necessary
to perform the calculation with the Green functions \cite{Abrikosov}.
For instance, one obtains
\begin{eqnarray}
&& \lim_{\delta \to +0} \sum_{n = -\infty}^{+\infty}
\mbox{Tr} [\ln {\cal G}^{-1} (i \omega_{n}, \mbox{\bf k})
e^{i \delta \omega_{n} \tau_{3}}] =
\lim_{\delta \to +0} \left\{ \sum_{n = -\infty}^{+\infty}
\mbox{Tr} [\ln {\cal G}^{-1} (i \omega_{n}, \mbox{\bf k})]
\cos \delta \omega_{n} + \right.
\nonumber\\
&&\left.
i \sum_{\omega_{n} > 0} \sin \delta \omega_{n}
\mbox{Tr} [(\ln {\cal G}^{-1} (i \omega_{n}, \mbox{\bf k}) -
\ln {\cal G}^{-1} (- i \omega_{n}, \mbox{\bf k})) \tau_{3}] \right\}
\nonumber\\
&& =\sum_{n = -\infty}^{+\infty}
\mbox{Tr} [\ln {\cal G}^{-1} (i \omega_{n}, \mbox{\bf k})] -
\frac{\xi(\mbox{\bf k})}{T},
\label{B4}
\end{eqnarray}
where
\begin{displaymath}
\ln\frac{ {\cal G}^{-1} (i \omega_{n}, \mbox{\bf k})}{i\omega_n}\simeq
\frac{- \tau_3\xi(\mbox{\bf k})+\tau_1\rho}{i \omega_{n}}, \quad
\omega_{n} \to \infty
\end{displaymath}
and
\begin{displaymath}
\sum_{\omega_{n} > 0} \frac{\sin \delta \omega_{n}}{\omega_{n}}
\simeq \frac{1}{2 \pi T} \int_{0}^{\infty} dx\,
\frac{\sin \delta x }{x} = \frac{1}{4 T}\, \mbox{sign}\, \delta.
\end{displaymath}

To calculate the sum in (\ref{B4}), one must first use
the identity $\mbox{Tr} \ln \hat A = \ln \det \hat A$, so that
(\ref{B1}) takes the form
\begin{equation}
\Omega_{\rm pot}(v, \mu, T, \rho) = v \left\{
\frac{\rho^{2}}{V} -  T \sum_{n = -\infty}^{+\infty}
\int \frac{d {\mbox{\bf k}}}{(2 \pi)^{2}}\,
\ln \frac{\det {\cal G}^{-1} (i \omega_{n}, \mbox{\bf k})}
{\det G_{0}^{-1} (i \omega_{n}, \mbox{\bf k})}
 -   \int \frac{d {\mbox{\bf k}}}{(2 \pi)^{2}}\,
[-\xi (\mbox{\bf k}) + \varepsilon (\mbox{\bf k})] \right\}.
\label{B5}
\end{equation}
Calculating the determinants of the Green functions (\ref{B2})
one obtains
\begin{equation}
\Omega_{\rm pot}(v, \mu, T, \rho) = v \left\{
\frac{\rho^{2}}{V} -  T \sum_{n = -\infty}^{+\infty}
\int \frac{d {\mbox{\bf k}}}{(2 \pi)^{2}}
\ln \frac{\omega_{n}^{2} + \xi^{2}(\mbox{\bf k}) + \rho^{2}}
{\omega_{n}^{2} + \varepsilon^{2}(\mbox{\bf k})}
-  \int \frac{d {\mbox{\bf k}}}{(2 \pi)^{2}}
[-\xi (\mbox{\bf k}) + \varepsilon (\mbox{\bf k})] \right\},
\label{B6}
\end{equation}
where the role of $G_{0} (i \omega_{n}, \mbox{\bf k})$ in the
regularization of $\Omega_{\rm pot}$ is now evident . The summation
in (\ref{B6}) can be done if one uses the representation
\begin{equation}
\ln \frac{\omega_{n}^{2} + a^{2}}{\omega_{n}^{2} + b^{2}} =
\int_{0}^{\infty} dx
\left(\frac{1}{\omega_{n}^{2} + a^{2} + x} -
      \frac{1}{\omega_{n}^{2} + b^{2} + x} \right),
\label{B7}
\end{equation}
and then
\begin{equation}
\sum_{k = 0}^{\infty}\frac{1}{(2k+1)^{2} + c^{2}} =
\frac{\pi}{4c} \tanh \frac{\pi c}{2}.
\label{B8}
\end{equation}
We find
\begin{equation}
\ln \frac{\omega_{n}^{2} + a^{2}}{\omega_{n}^{2} + b^{2}} =
\int_{0}^{\infty} dx   \left(
\frac{1}{2 \sqrt{b^{2} + x}} \tanh \frac{\sqrt{b^{2} + x}}{2 T} -
\frac{1}{2 \sqrt{a^{2} + x}} \tanh \frac{\sqrt{a^{2} + x}}{2 T}
\right).                     \label{B9}
\end{equation}
Integrating (\ref{B9}) over $x$, one thus obtains
\begin{equation}
T \sum_{n = -\infty}^{+\infty}
\int \frac{d {\mbox{\bf k}}}{(2 \pi)^{2}}
\ln \frac{\omega_{n}^{2} + \xi^{2}(\mbox{\bf k}) + \rho^{2}}
{\omega_{n}^{2} + \varepsilon^{2}(\mbox{\bf k})} =
2 T  \int \frac{d {\mbox{\bf k}}}{(2 \pi)^{2}}
\ln \frac{\cosh [\sqrt{\xi^{2}(\mbox{\bf k}) + \rho^{2}}/2T ]}
{\cosh [\varepsilon(\mbox{\bf k})/2T ]}.       \label{B10}
\end{equation}
Finally, substituting (\ref{B10}) into (\ref{B6}),
\begin{equation}
\Omega _{\rm pot}(v, \mu, T, \rho) = {v} \left\{
\frac{\rho^2}{V} -
\int \frac{d \mbox{\bf k}}{(2 \pi)^{2}}
\left[2T
\ln\frac{\cosh[{\sqrt{\xi^2(\mbox{\bf k}) + \rho^2}/{2T}}]}
{\cosh[{\varepsilon(\mbox{\bf k})/{2T}}]} -
[\xi(\mbox{\bf k})- \varepsilon(\mbox{\bf k})] \right] \right\}.
                                         \label{B11}
\end{equation}
It is easy to show that at $T=0$, the expression (\ref{B11}) reduces to
that obtained in Ref. \cite{GGL}.

Finally, we give formulas for the summation over the Matsubara frequencies
used in Secs.~III and VI:

\begin{eqnarray}
& & T \sum\limits_{n=-\infty}^\infty\mbox{Tr}
[{\cal G}(i\omega_{n},\mbox{\bf k}) \tau_3
{\cal G}(i \omega_{n}, \mbox{\bf k}) \tau_3] =
2T \sum\limits_{n=-\infty}^\infty
\frac{\xi^{2}(\mbox{\bf k})-\rho^2 - \omega_ {n}^{2}}
{[\omega_{n}^{2} + \xi^{2}(\mbox{\bf k}) + \rho^{2}]^{2}}
\nonumber                \\
& & =  - \frac{\rho^2}{[\xi^2(\mbox{\bf k}) + \rho^2]^{3/2}}
\tanh \frac{\sqrt{\xi^2(\mbox{\bf k}) +\rho^2}}{2T} -
\frac{\xi^2(\mbox{\bf k})}{2T[\xi^2(\mbox{\bf k})+\rho^2]}
\frac{1}{\cosh^2 \frac{\ds \sqrt{\xi^2(\mbox{\bf k})+\rho^2}}{\ds 2T}},
\label{B12}
\end{eqnarray}

\begin{eqnarray}
& & T \sum \limits_{n=-\infty}^\infty \mbox{Tr}
[{\cal G} (i\omega_{n},\mbox{\bf k}) \hat I
{\cal G}(i \omega_{n}, \mbox{\bf k}) \hat I] = 2T
\sum \limits_{n=-\infty}^\infty \frac{\xi^{2}(\mbox{\bf k}) + \rho^2 -
\omega_ {n}^{2}}{[\omega_{n}^{2} +
\xi^{2}(\mbox{\bf k}) + \rho^{2}]^{2}}
\nonumber              \\
& & =  - \frac{1}{2T} \frac{1} {\cosh^{2}
\frac{\ds \sqrt{\xi^{2}(\mbox{\bf k}) + \rho^{2}}}{\ds 2T}},
                     \label{B13}
\end{eqnarray}
where the Green function ${\cal G}(i \omega_{n}, \mbox{\bf k})$ is given by
(\ref{A2}). Both formulas can easily be calculated using
Eq. (\ref{B8}) and its derivative with respect to $c$.

\newpage

\begin{center}
FIGURE CAPTION
\end{center}

\vspace{1cm}
\noindent
FIG.~1.

$T_{\rm BKT}$ and $T_{\rho}$ versus the noninteracting fermion density.
Dots represent the function $\rho(\epsilon_{F})$ at $T = T_{\rm BKT}$.
The regions of normal phase (NP), pseudogap phase (PP), and BKT
phase are indicated.

\vspace{1cm}

\noindent
FIG.~2.

$\mu(T)$ for various values of $\epsilon_{F}/|\varepsilon_{b}|$:
1) 0.05; 2) 0.2; 3) 0.45; 4) 0.6; 5) 1; 6) 2; 7) 5.
(For $\mu > 0$ and $\mu < 0$ the chemical potential was scaled to
$\epsilon_{F}$ and $|\varepsilon_{b}|$, respectively.) The thick lines
delimit regions of BKT, pseugogap (PP), and normal (NP) phases.

\vspace{1cm}

FIG.~3.

\noindent
$2 \Delta/T_{\rm BKT}$ and $2 \Delta/T_{\rho}$ versus the
non-interacting fermion density.

\vspace{1cm}

FIG.~4.

\noindent
Phase diagram of the 2D metal at low concentrations.
The dotted line corresponds to $\mu =0$, and the temperature $T_{P}$
separates abnormal normal phase (ANP) from normal phase.
The critical temperature $T_{\rm BKT}$ is not shown.

\vspace{1cm}

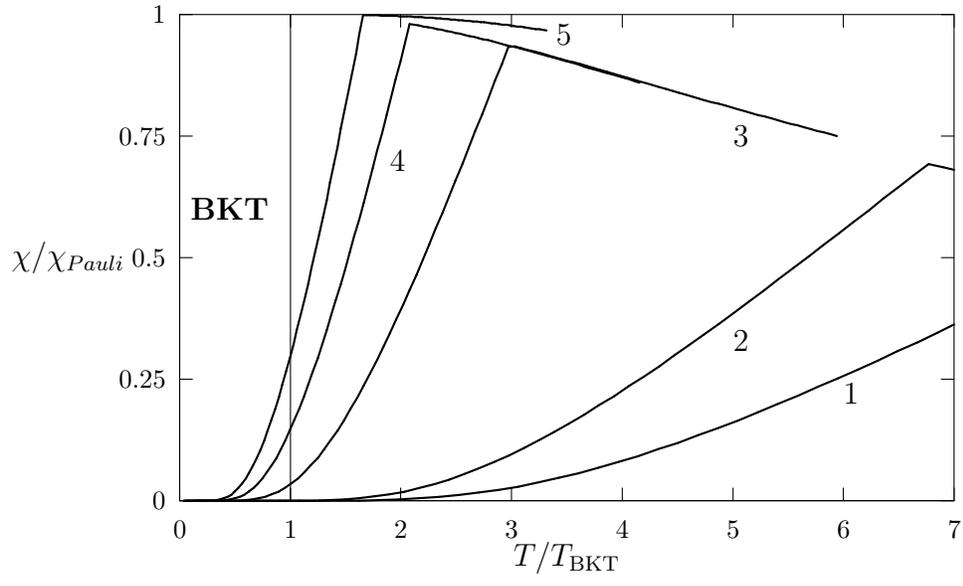
FIG.~5.

\noindent
$\chi(T)$ for various values of
$\epsilon_{F}/|\varepsilon_{b}|$: 1) 0.6; 2) 1;
3) 5; 4) 10; 5) 30.

%\iffalse
\newpage

\begin{figure}
% GNUPLOT: LaTeX picture with emtex specials
\setlength{\unitlength}{0.240900pt}
\ifx\plotpoint\undefined\newsavebox{\plotpoint}\fi
\sbox{\plotpoint}{\rule[-0.200pt]{0.400pt}{0.400pt}}%
\special{em:linewidth 0.4pt}%


\caption{}
\end{figure}

\newpage

\begin{figure}
% GNUPLOT: LaTeX picture with emtex specials
\setlength{\unitlength}{0.240900pt}
\ifx\plotpoint\undefined\newsavebox{\plotpoint}\fi
\sbox{\plotpoint}{\rule[-0.200pt]{0.400pt}{0.400pt}}%
\special{em:linewidth 0.4pt}%
%

\caption{}
\end{figure}

\newpage
\begin{figure}
% GNUPLOT: LaTeX picture with emtex specials
\setlength{\unitlength}{0.240900pt}
\ifx\plotpoint\undefined\newsavebox{\plotpoint}\fi
\sbox{\plotpoint}{\rule[-0.200pt]{0.400pt}{0.400pt}}%
\special{em:linewidth 0.4pt}%
%

\caption{}
\end{figure}
\newpage

\begin{figure}
% GNUPLOT: LaTeX picture with emtex specials
\setlength{\unitlength}{0.240900pt}
\ifx\plotpoint\undefined\newsavebox{\plotpoint}\fi
\sbox{\plotpoint}{\rule[-0.200pt]{0.400pt}{0.400pt}}%
\special{em:linewidth 0.4pt}%
%

\caption{}
\end{figure}

%\fi

\end{document}